# MODELING THE OPERATION OF AN AUTONOMOUS PROFILOGRAPH AS A DYNAMIC SYSTEM WHEN MEASURING OCEANOLOGICAL FIELDS

***L.A. Krasnodubets, A.N. Grekov****

Sevastopol State University
Sevastopol, St. Student, 33
E-mail: lakrasno@gmail.com
*Institute of Natural and Technical Systems
Russia, Sevastopol, 28 Lenin Street
E-mail: oceanmhi@ya.ru

*On the basis of the proposed model of a diving buoy with adjustable buoyancy, a model of a double-loop adaptive control system for the speed modes of an autonomous profiler has been developed, depending on the gradients of the measured parameters. The results of modeling are presented.*

**Introduction.** Current measurements of the vertical structure of the ocean environment include the registration of field data from various sensors with their subsequent processing and dissemination within the framework of the global observing system GOOS [1]. As the basic measuring component of this system, ARGO class autonomous diving buoys with adjustable buoyancy [2] were selected, which, acting as mobile data collection platforms, record temperature and conductivity profiles every 10 days for 2–4 years and transmit the data to processing center via satellite communication channel. According to [3], as of 03.09.2015, there were 3881 Argo-class buoys in the active state. The main disadvantage of such buoys, which are essentially marine robotic profilers, is their low sinking speed, which approximately corresponds to 0.1 m/s [4]. As shown in [5], probing with such a speed practically eliminates dynamic distortions when measuring temperature using temperature sensors from SBE [6], the time constant of which is estimated by the value $\theta_t = 0{,}065\,c$. However, sounding at the indicated speed mode to a depth of 2000 m requires a significant amount of time (~ 6 hours). This can lead to distorted measurements due to the temporal and spatial variability of the ocean environment. The practice of oceanological research has shown that the range of values of the measured parameters with large gradients extends on average to depths of 500 m. For this reason, there is no need to carry out sounding at a low speed of the entire profile, and in order to reduce the time and, accordingly, save money for the experiment, provide a flexible change speed depending on the gradients of the measured parameters using a specially developed control system.

The purpose of this article is to develop and study by the method of computer simulation an adaptive system for automatic control of the speed modes of movement of a diving buoy, depending on the gradients of the measured parameters of the ocean environment. At the same time, the WOCE recommendations [7] regarding the measurement error of practical salinity, temperature and pressure, which should not exceed $\Delta Sp = 0{,}002$, $\Delta t = 0{,}002$ °C, $\Delta P = 3$ dbar ( $\Delta H = 3$ м ) in terms of salinity, temperature and pressure, were taken as a criterion for choosing a speed mode [7].

**Model of a diving buoy with adjustable buoyancy as a control object.** The movement of a diving buoy as a cylindrical float device is investigated under the action of a driving force (a component of the Archimedes force), which is formed by a buoyancy control device built on the basis of a special camera, which is an element of the buoy hull and is capable of changing its linear size – length. The investigated movement of the buoy in the aquatic environment will be considered in the vertical plane without taking into account its horizontal component. This is justified by the fact that when submerging or ascending, the buoy freely displaces (drifts) in a horizontal direction parallel to the surface, together with the masses of the surrounding water.

The vertical movement of a diving buoy, for example, when diving along an axis $Z$, occurs under the influence of three main forces. This is the force of gravity $P$, the force of hydrodynamic resistance $F_c$ and the total pressure from the environment, the main component of which generates a buoyant (Archimedean) force $F_в$, which includes the control force $F_u$.

Other components of the ambient pressure are caused by various hydrodynamic processes, for example, the passage of waves, as well as the inhomogeneous distribution of the chemical composition and physical characteristics of the aquatic environment over depth. The latter significantly affect the value of the density of seawater $\rho_в$, on which the magnitude of the buoyancy force depends. The volume of the buoy hull submerged in water $V_б(t)$ also includes the volume of the buoyancy chamber $V_{кп}(t)$, the value of which can be purposefully changed. The hydrodynamic drag force $F_c$ is the result of pressure disturbances caused by the movement of the buoy in a water medium with different densities, as well as friction stresses. Under the known assumption [8] that the hydrodynamic drag force is proportional to the square of the movement speed a diving buoy in a water environment and has a sign opposite to the direction of the vector of this velocity, the expression for calculating this force has the form

$$F_c = \frac{C_x S_m \rho_в}{2}(v - v_l)\left|(v - v_l)\right|, \quad (1)$$

where $C_x$ is the drag coefficient; $S_m$ - the area of the maximum cross-section of the diving buoy hull; $v$ - speed of movement of a diving buoy in sea water; $v_l$ - local water velocity; $g_э$ - effective acceleration of gravity; $\rho_в$ is the density of sea water.

The control force arises as a result of an increase in the volume of the body part due to a change in the volume of the buoyancy chamber $V_{кп}(t)$ in the range from 0 to $V_{кп}{}^{\max}$, taking the shape of a cylinder with a diameter approximately equal to the diameter of the main body of the buoy. Therefore, we can assume that the base area of this cylinder is equal to the cross-sectional area of the buoy body $S_c$, and its current length $l_c(t)$ is equal to the linear displacement of the pump piston $l(t)$, since the pumped the hydraulic fluid is practically not compressed.In this case, the current value of the volume of the buoyancy chamber is expressed as follows

$$V_{кп}(t) = l(t) S_c,$$

where $0 \le l(t) \le l^{\max}$ and accordingly

$$0 \le V_{кп}(t) \le V_{кп}{}^{\max}. \quad (2)$$

Taking into account (2), let us make the assumption that the buoy design is such that it will have neutral buoyancy if $V_{кп}(t) = \frac{V^{\max}{}_{кп}}{2}$. In this case, the inequality for $l(t)$ from (2) takes the form

$$-\frac{l^{\max}}{2} \le l(t) \le \frac{l^{\max}}{2}. \quad (3)$$

In what follows, we will assume that the change in the volume of the buoyancy chamber is carried out by a special electrohydraulic drive with feedback. In this case, the movement of the pump piston occurs with the help of a screw (threaded) shaft, driven by a DC electric motor together with a gearbox, and is controlled by negative feedback. The equation of the electro-hydraulic device that changes the buoyancy of the buoy and, thereby, creates a control force $F_u$, can be written in the form

$$T_e \frac{dl(t)}{dt} + l(t) = k_e u(t), \quad (4)$$

where $T_e$ is the time constant of the electrohydraulic device; $l(t)$ is the linear displacement of the pump piston; $k_e$ is the transmission coefficient of the electrohydraulic device; $u(t)$ is the control voltage.

By projecting the forces acting on the diving buoy on the vertical axis of the associated coordinate system, taking into account Newton's second law, we can obtain the following equation describing the dynamics of the diving buoy's motion:

$$\ddot{z}(t) = -a|\dot{z}(t)|\dot{z}(t) - bu(t) + f_p(z)$$
$$t \in [t_0, t_f], \qquad (5)$$

where $z(t)$ is the immersion depth.

$$a = \frac{C_X S}{2V_б^0},\ b = \frac{Sgk_p}{V_б^0},$$

$$f_p(z) = \left[1 - \frac{\Delta\rho_в(z)}{\rho_в(z)}\right] g. \qquad (6)$$

Nonlinear inhomogeneous differential equation (5) with initial conditions

$$t_0 = 0; \quad z(t_0) = z_0; \quad \dot{z}(t_0) = \dot{z}_0$$

describes the processes of movement (immersion or ascent) in the vertical plane of a diving buoy, which is under the influence of three main forces:

– the forces of resistance to motion, represented by the first term in the right-hand side of equation (5);

– buoyancy force, represented by the second term on the right side of equation (5) and performing the control function;

– the force represented by the third term on the right-hand side of equation (5), which is the resultant of gravity and a component of the buoyancy force resulting from a change in the density of seawater. During the dive, this resultant acts as a driving force. When ascending, the buoyant force acts as a driving force, which fully compensates for the force of gravity and provides a control effect.

A feature of the model of motion of a diving buoy in the form (5) from a similar model in [9] is the reduction of perturbations caused by changes in the stratification (vertical density profile) of the marine environment to the resultant force $f_p(z)$, which depends not only on depth, but also, strictly speaking, However, given the small variability in time of this characteristic, the force $f_p(z)$ can be considered as a function of only the depth $z$.

In this case, the parameters of model (5), respectively $a$ and $b$ remain constant. This greatly facilitates the modeling of the processes of controlled movement of a diving buoy in conditions of strong stratification of the environment), as well as the calculation of gradients of its density and related parameters.

It should be noted that for equation (5), along with the initial conditions, one should also take into account the inequalities

$$-\frac{u^{\max}}{2} \le u(t) \le \frac{u^{\max}}{2},$$

where $u^{\max} = \frac{l^{\max}}{k_p}$, defining constraints on the control function $u(t)$.

The block diagram of modeling a diving buoy as a control object is shown in fig. 1 as an S-model for modeling in a visual modeling environment *Simulink.*

The diagram shows the main modules of the mechatronic buoyancy control system: *Diving buoy* – diving buoy, which is a nonlinear dynamic system of the second order, the variables of which are the depth and speed of sinking (ascent); *Hydraulic drive* – an electro-hydraulic drive, which is a dynamic system with a constraint, the description of which is made in the form of a first-order transfer function and a non-linear element; *Profiler model* – a model containing arrays of measurements taking into account the time constants of the sensors of thermohaline parameters, as well as the density of seawater as a function of depth.

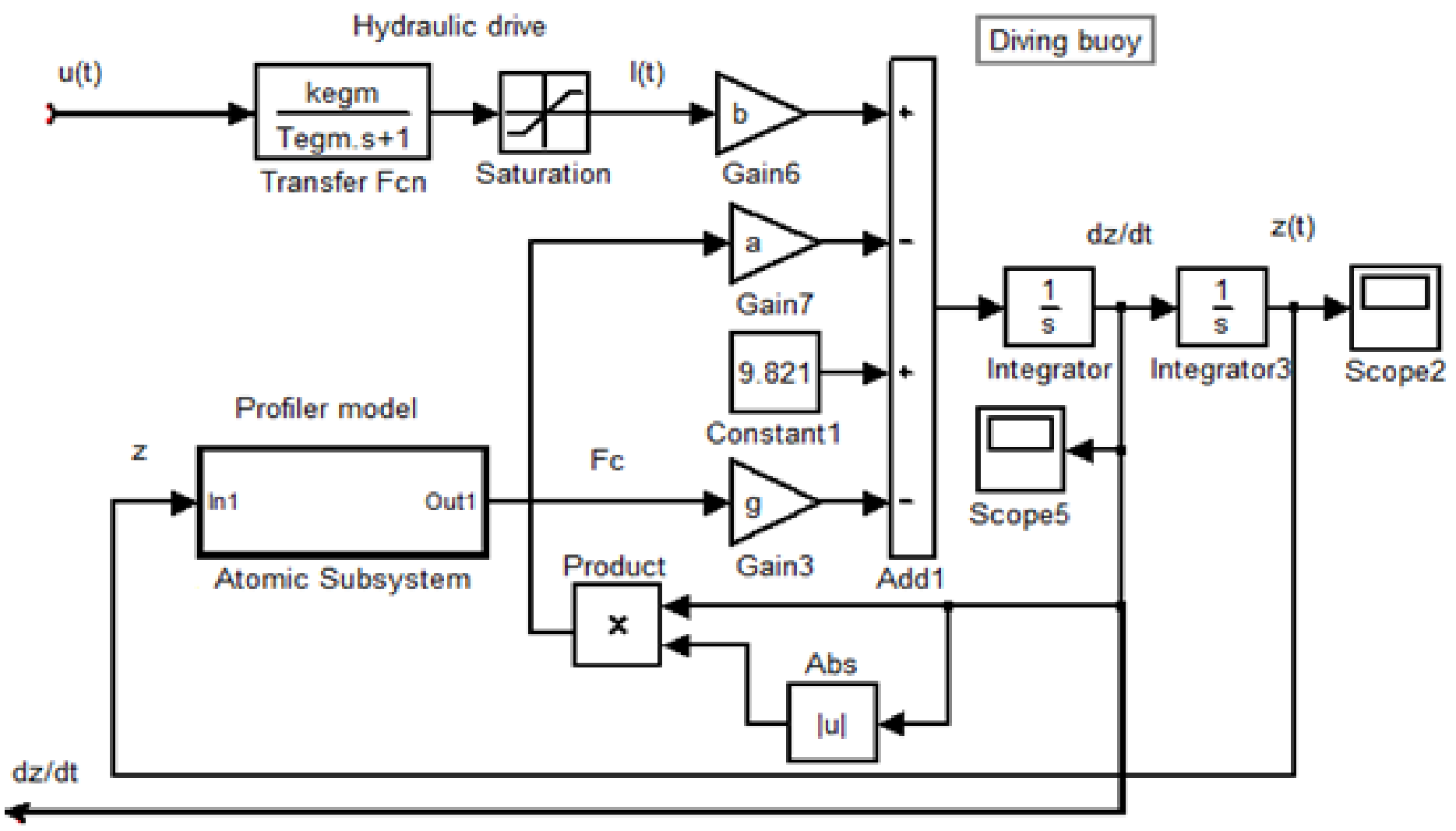

Fig. 1. Block diagram of the simulation of a mechatronic system - a diving buoy with a drive

**An adaptive regulator of the automatic control system for the speed modes of the vertical movement of an autonomous profiler.** The control law for the adaptive regulator designed to control the speed modes of the profiler, we construct, following the method [10]. Introduce in equation (5) the change of variable $\dot z = v(t)$. We have

$$\dot v(t) = -a|v(t)|v(t) - bu(t) + f_p(z)\,. \qquad (7)$$

Next, we set the task: find the control function $u(v,t)$ in the form of feedback, which will ensure the transfer of the control system from the initial speed mode $t = t_0;\ v(t_0) = v_0$ to the new speed mode $t = t_f;\ v(t_f) = \bar v$.

In this case, we require that the phase trajectories of the process $v(t) \to \bar v$ pass in a small neighborhood of the phase trajectories of the reference process $v^*(t) \to \bar v$, specified by the reference model in the form of a differential equation

$$\dot v^*(t) + \alpha v^*(t) = \alpha \bar v\,,\ t \in [t_0, t_f]\,, \qquad (8)$$

where the parameter $\alpha > 0$, which determines the dynamic and static properties of the reference process, must be selected. In this case, as a measure of the residual of the processes $v^*(t) \to \bar v$ and $v(t) \to \bar v$ we take the objective function

$$G(u) = 0{,}5[\dot v^* - \dot v(u)]^2\,. \qquad (9)$$

The desired control function $u(t)$ we find from the condition $G(u) = \min$ by the gradient descent method [11]), according to which

$$\frac{du}{dt} = -\lambda \frac{\partial G(u)}{\partial u}\,,\ \lambda = const\,. \qquad (10)$$

Calculating the derivative (9), we bring equation (10) taking into account (7) and (8) to the form

$$\frac{du}{dt} = \lambda b[\alpha(\bar v - v^*) - \dot v]\,. \qquad (11)$$

Further, integrating both sides of (11) with zero initial conditions, one can find an expression for the sought law of adaptive control with the reference model

$$u(t) = \lambda b[\alpha \int_0^t (\bar v - v^*)dt - v]\,. \qquad (12)$$

When generating the control signal in accordance with the found control law, the following signals should be included in expression (12): the signal for setting the speed mode $\bar v$, the signal from the output of the reference model $v^*$ and the negative feedback signal $v^* = v$. adaptive control with an implicit reference model, which will take the following form

$$u(t) = \lambda b[\alpha \int_0^t (\bar v - v)dt - v]\,. \qquad (13)$$

The structural diagram of an adaptive controller designed to implement the control law (13) is presented in the form of an S-model in fig. 2.

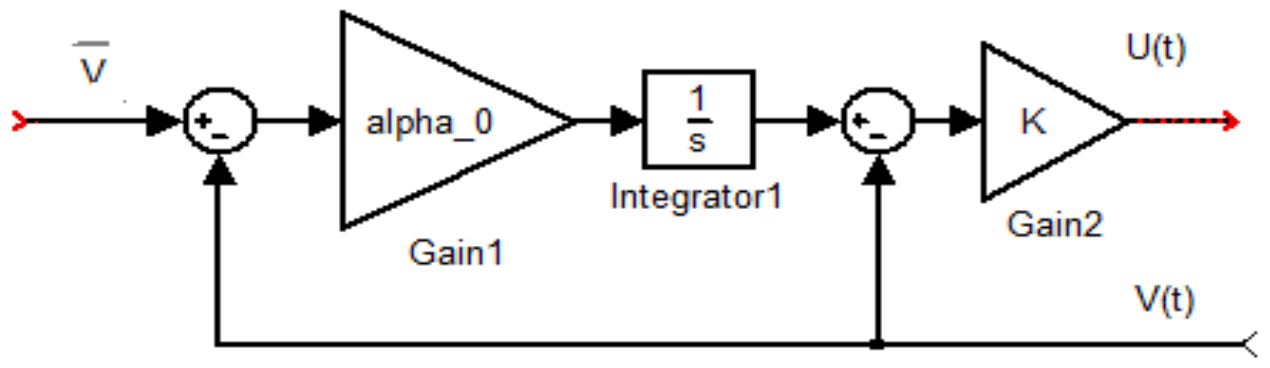

Fig. 2. Block diagram of the speed regulator

**Modeling of high-speed modes of an autonomous profiler as a microcontroller system.** Since the design of the regulator together with the mechatronic buoyancy control system assumes a technical design based on a microcontroller system, a discrete-time simulation scheme was developed to study such a system, which was obtained by transforming the continuous models shown in fig. 1 and fig. 2, to the discrete methods of Tustin [12].

The developed model is presented in the form of a structural diagram (S-model) shown in fig. 3, and a file with a control program (m-file), oriented for execution in the Matlab & Simulink visual modeling environment [13].

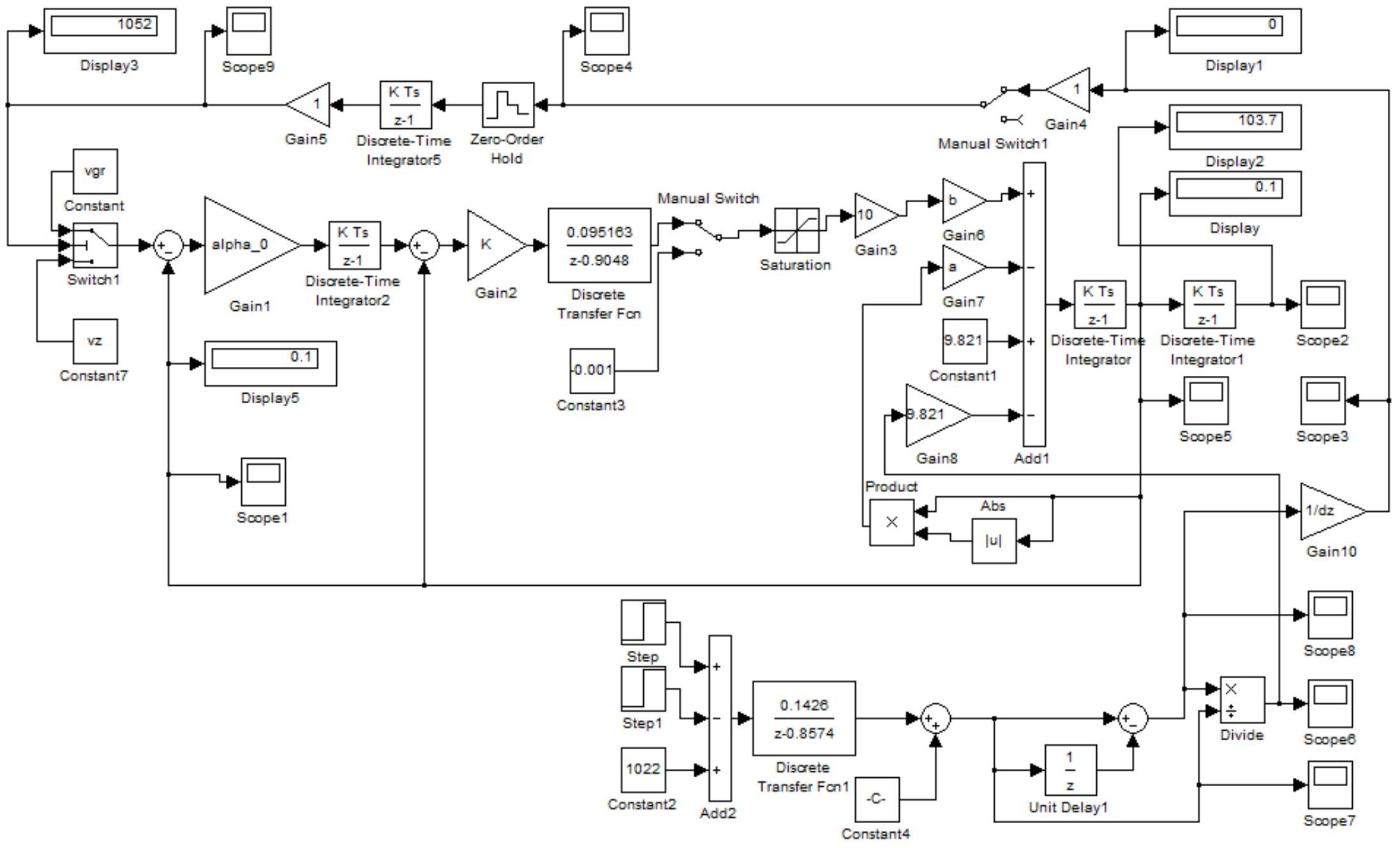

Fig. 3. S-model for the study of the adaptive speed control system autonomous profiler

The control program is designed to provide simulation processes, as well as to calculate the parameters of the controller and in accordance with the methodology [14] and perform calculations associated with the transformation of the original continuous model to a discrete one. The proposed S-model operates in discrete time with a processor clock frequency and has a block-modular structure including the following submodules: control object (diving buoy with adjustable buoyancy, structurally made in the form of a mechatronic system); adaptive regulator for automatic stabilization of the set speed of descent (ascent); a submodule that simulates the vertical profile of the oceanic medium stratification as a function of depth; a submodule that simulates an information-measuring channel taking into account its dynamic properties; elements that form the setting and disturbing influences; elements of registration and visualization of simulated processes, as well as switches.

The control system for speed modes of an autonomous profiler has two circuits. The first control loop is designed to automatically stabilize the speed of vertical movement, which is set by the value of a constant $vz = 1$ ("cruising" mode of movement with a speed $v = 1$ m/s)) or a constant $vgr = 0{,}1$ (slow "measuring" mode of movement with a speed $v = 0{,}1$ m/c, providing a low level of dynamic distortions caused by the inertia of the sensors) The second circuit calculates and records the relative changes in the density of the surrounding water, which are transmitted to the dynamic model of the diving buoy and the device (automatic switch) for selecting the speed mode of the profiler movement. A sharp deviation of the density of seawater from the initial average statistical value of 1022 $кг/м^3$, recorded by the second control loop, leads to an automatic change in the speed mode of the profiler from "cruising" to "measuring" by supplying a reference signal $vgr = 0{,}1$ to the first loop (speed stabilization loop), which provides a decrease in speed from $v = 1$ m/s to $v = 0{,}1$ m/s.

In this case, the further movement of the profiler with the speed set for the "measurement" mode continues under the control of the stabilization system, which fends off parametric and other external disturbances for some time (set in the simulation settings), and then its movement switches back to the "cruise" mode. Thus, the automatic control system for the movement of the autonomous profiler sets a "measuring" speed mode whenever the current measurement data begins to change abruptly relative to the previous ones. In this case, the “measuring” mode switches back to the “cruising” mode when the data loses its variability at a certain predetermined interval.

The results of modeling the dynamics of the change in speed modes under the influence of gradients in the density of seawater are shown in Fig. 4. It is assumed that such gradients in real conditions can be indirectly determined from temperature gradients.

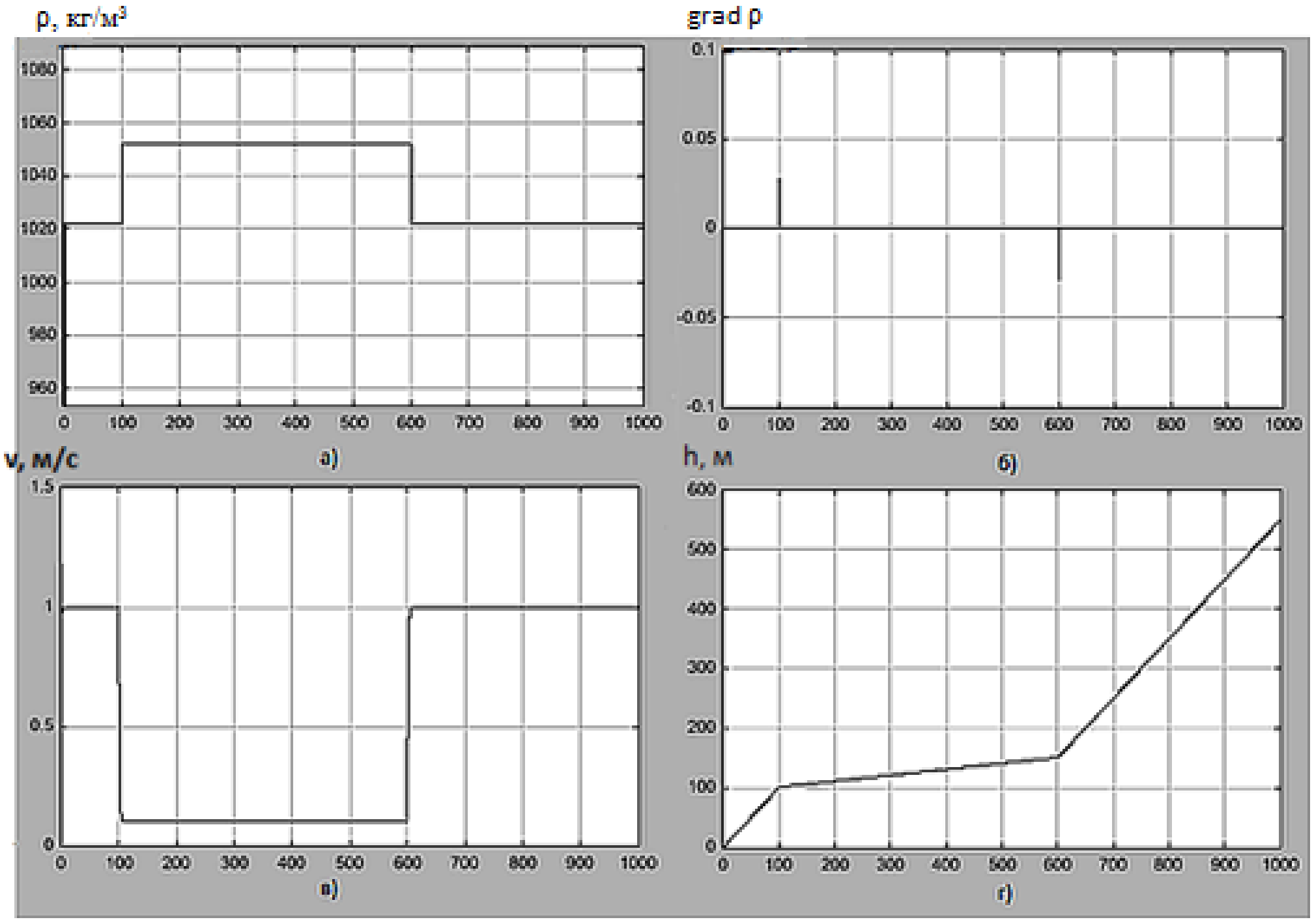

Fig. 4. Graphs of processes in a dynamic measurement system when changing the speed dive mode "cruising" to "measuring":
a) the density of sea water; b) water density gradients; c) sinking speed; d) depth

**Conclusion.** Controlling the processes of profile measurements by changing the speed modes of movement of an autonomous profiler in order to reduce dynamic distortions in the measuring channels is an effective way to improve the quality and representativeness of information on the vertical stratification of the oceanic environment and, in particular, to detect its fine structure. At the same time, by reducing the time for profiling, it is possible to reduce the cost of conducting experimental studies and reduce the volume of memory devices intended for storing time series of data, which in this case do not require further processing and can be used in real time.